# Energy Efficient Cloud-Fog Architecture

Hatem A. Alharbi, Taisir E.H. Elgorashi and Jaafar M.H. Elmirghani
School of Electronic and Electrical Engineering, University of Leeds, LS2 9JT, United Kingdom

*Abstract*— The advancements of cloud computing came as a radical transformation in the way Information and Communication Technology (ICT) services are deployed and maintained. Cloud computing provides ubiquitous on-demand access to an Internet-based pool of processing, storage, and communication resources offered to a large set of geographically distributed users. As the cloud computing infrastructure grows and demand increases, the need for a new breed of on-demand computing that can efficiently maintain Quality of Service (QoS) requirements has increased. Fog computing was proposed to address the limitations of cloud computing, in terms of delay and high bandwidth requirements, by extending the on-demand resources of clouds to the edge of the network bringing them closer to the users. The massive growth and wide use of cloud-fog services have created serious power consumption concerns. This article delves into the energy consumption of cloud-fog services by raising headline questions related to; how significant the problem itself is, how different conditions/scenarios affect the energy consumption of the architecture, and how to orchestrate the use of the architecture in an energy-efficient manner. We start by summarizing the cloud-fog architecture including different communication and computing layers. Additionally, we give a brief overview of the role of Virtual Machine (VM) placement in optimally using cloud-fog resources in a dynamic manner. Then, we present the problem of energy efficient VMs placement and provide numerical results.

## I. INTRODUCTION

The significant impact of Information and Communications Technologies (ICT) services on people daily lives led to an increasing perception that cloud computing is the 5th utility after water, electricity, gas, and telephony. Cloud Computing has dominated the ICT industry by providing efficient resource sharing solutions where an Internet-based pool of network, storage and computational resources is made available to simultaneously serve a large number of geographically distributed users. Cloud computing can provide the bandwidth, memory, and processing capability needed to serve Big Data, Internet of things (IoT) and Artificial Intelligence (AI) applications such as image recognition, video analytics, augmented and virtual reality.

According to Cisco, in 2017, the global cloud computing traffic was 56% of the Internet traffic. Further growth is projected as global cloud computing traffic is expected to hit 72% of the Internet traffic in 2022 [1]. This proliferation in data volume and processing requirements increases the need for a new breed of on-demand computing placement and administration. Fog computing is proposed by academia and industry to bring cloud services closer to users. Fog computing complements the clouds by extending processing, networking and storage resources to the edge of the network. Offloading tasks to fog nodes is proposed to provide low operating cost, low latency, preserve network bandwidth, provide real-time analytics and interaction, improve security and improve Quality-of-Service (QoS) and Quality-of-Experience (QoE) for different computing services [2].

As a result of the significant growth in demand for ICT services, energy efficiency has been recognized in the last few years as one of the core requirements needed to develop a sustainable ICT infrastructure. By 2025, the ICT industry is projected to consume 20% of the global electricity demand [3]. Virtualization has been proposed as an enabler for energy efficient cloud-fog services through the consolidation of resources [4]. In this article, we present a comprehensive review of VM placement over cloud-fog architecture. Cloud and fog processing employs Virtual Machines (VMs) for efficient resource utilization. Virtualization abstracts the server resources including the CPU, RAM, hard disk and I/O network to create an isolated virtual entity that can run its operating system and applications. The existence of such a virtual environment allows the scaling up and down of server resources in a dynamic manner based on the variation in user demands [5].

This article reviews the existing work on energy efficient cloud-fog architectures. To the best of our knowledge, this is the first review of this topic. We start by outlining the main communication network layers that are essential in the realization of an energy efficient cloud-fog architecture. We then present the cloud-fog computing layers made up of infrastructure, and service layers as well as the issues related to service management and orchestration. This leads us to identify the main factors that affect the design of an energy efficient cloud-fog architecture.

## II. CLOUD-FOG ARCHITECTURE

In this section, we provide background on the key aspects that are necessary to realize the cloud-fog network architecture. We follow that by an overview of cloud and fog computing architectures and the role of VMs in providing dynamicity in these architectures.

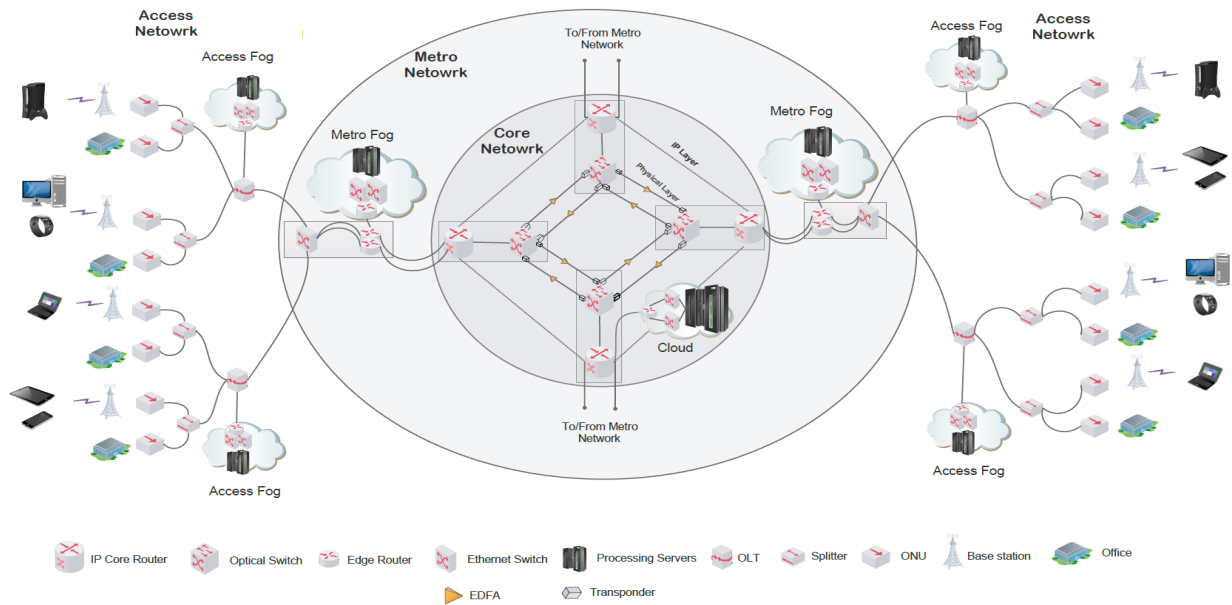

Figure 1: Cloud-Fog Architecture [5].

*A. Communication Networks*

The traditional Internet protocol (IP) network structure as used by ISPs, can logically be split into three main layers; the core network, the metro network and the access network. Fig. 1 illustrates this three layers architecture. The core network represents the backbone infrastructure of any telecom network as it interconnects major regions/cities. IP over wavelength division multiplexing (WDM) technology is widely deployed in the core network due to its ability to provide high capacity, scalability and transfer speed. Based on communication networks hierarchy, each core node is connected to a metro network, which covers a metropolitan area. Metro Ethernet is the dominant technology used in enterprise metro network. It provides direct connectivity between residential users in access networks and core network node. The access network represents the last mile of the telecom network, connecting telecom offices and end-users. Passive optical networks (PONs) are the main technology deployed broadly in the access network.

*B. Cloud and Fog Computing Layers*

This section provides an overview of cloud and fog computing architectures and the role of VMs in providing dynamicity in these architectures. Related work from the literature on optimizing VMs placement in cloud-fog architectures is also reviewed.

1) *Cloud Computing Layer:*

Cloud computing provides ubiquitous on-demand access to an Internet-based pool of computing, storage, and communication resources. These resources can be provided to a large set of geographically distributed users. Typically, the cloud interface resides within a single window in the users' Internet browser. According to the American National Institute of Standards and Technology (NIST) [6], the cloud computing model is essentially composed of five features; on-demand self-service, online availability through any platform (e.g. smartphones, laptops), a pool of computing resources available to end-users through a multitenancy architecture, and elastic and measurable service.

**Cloud computing service models:** Cloud computing service models can be classified into three categories [6]; software as a service (SaaS), platform as a service (PaaS) and infrastructure as a service (IaaS). SaaS delivers an application to the end-users through cloud datacenters that can be accessed through users' web interfaces (e.g. websites, Health care systems, Geographic Information Systems (GIS), Microsoft office). PaaS offers end-users the environment needed to create applications on cloud infrastructure using a set of online tools (e.g. programming languages) to provide facilities for designing, developing, testing and deployment of applications. Examples of this model include Google App Engine and Microsoft Azure. Further abstracted computing resources are available to end-users through the IaaS model. In addition to the deployed applications, the end-user has control over the operating system, middleware (for enabling communication between two applications), runtime and data, in addition, to directly using networking, storage and compute resources, which are usually made available on a subscription basis. Amazon Elastic Compute Cloud (Amazon EC2) is a widely known solution based on the IaaS service model.

2) *Migrating to Fog:*

The concept of fog computing was introduced by Cisco in 2014 to bring cloud services closer to the users. According to the OpenFog Consortium [7], the fog computing architecture complements the clouds by extending processing, networking and storage resources to the edge of the network. The advent of fog computing together with cloud computing has introduced a promising paradigm shift. The research efforts studying fog computing have mainly focused on illustrating its potential advantages over cloud computing. Offloading tasks to fog nodes is proposed to provide real-time analytics, low operating expense, low latency, privacy, processing sensitive data locally instead of processing at distant data centres; together with improved QoS and QoE for different computing services.

**Traditional datacentre Architecture:** Cloud and fog datacenters typically consist of a wide range of servers arranged in multiple racks and a Local Area Network (LAN) made up typically of two switch layers (a spine and leaf architecture) used to connect racks to each other in order to enable inter rack communication. An aggregation router is used to connect the datacenter to the outside world, i.e. to users and other datacenters (inter-datacenters communication). The architecture of conventional cloud datacenters and disaggregated cloud data centres is discussed in [8].

3) *Machine Virtualization*

Virtualization is employed in the cloud-fog architecture to satisfy the users need to rapidly grow/shrink the usage of the datacenter physical resources. Here the datacentre physical resources are abstracted into several logical entities called VMs [4]. Each VM is allocated its resources of CPU, memory, network bandwidth and storage to run an application logically isolated from other applications running in other VMs.

**VM orchestration:** In a virtualization environment, VMs are orchestrated by the VM monitor (VMM) or hypervisor. The hypervisor is the virtualization software in a system that creates and runs one or multiple VMs (guest machines) over physical hardware (host machine), which is typically a server. The hypervisor has the authorization to control all the VM's operating system and its hosted application as well as server resources.

**Consolidation**: Using VMs, multiple applications can be consolidated into a fewer number of servers by allowing multiple heterogeneous virtual entities, each serving a different client, to coexist on a shared physical resource (server(s)) owned and operated by an infrastructure/service provider. Each physical server can host up to hundreds of VMs and each VM-hosted application allows multiple tenants or users to share a single application as it runs on a dedicated environment [9]. These virtual entities are created and torn down on demand to cater for the needs of the cloud clients, allowing for scalable growth and efficient use of resources through consolidating virtual entities in fewer physical resources. Fig. 2 illustrates the VM consolidation concept.

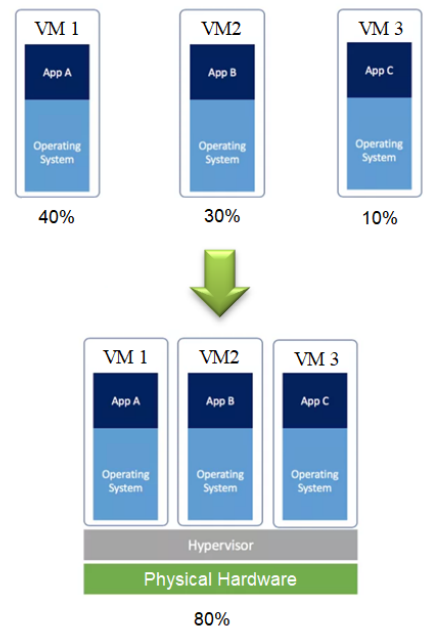

Figure 2: VMs consolidation.

**VM replication and migration:** Further dynamism in resource management can be achieved by placing or relocating VMs within or across distributed datacenters through either replication or migration. VM placement is a key operation in cloud and fog computing infrastructure management, where the most suitable server is found to host the VM based on workload balancing, datacentre maintenance, failover recovery or energy efficiency. Consider an example, where a content provider (CP) wants to run an application on a cloud architecture. At time $t = 1$, the application is unpopular and a single VM is enough to serve all users. At time $t = 2$, the application becomes popular, so, a single machine is no longer enough to serve all the users. Thus, an estimation can be made to find how many VM replicas are required to run the application by taking into consideration the server resources and the number of concurrent users accessing the application at that time. An elastic cloud and fog architecture should instantly react to the increased load and serve all the users of the application responsively.

**VM Categories**: Based on computation and network bandwidth requirements, cloud and fog applications can be classified into three categories [8]; 1) CPU-intensive applications; only require computation resources and produce a low data rate over the communication network. e.g. high-performance computing applications, 2) data-intensive applications; require fewer computation resources and produce high traffic between the communicating nodes. e.g. video applications. 3) balanced applications; both computation and communication network resources are required. e.g. GISs applications.

From a CPU perspective, studies in the literature have shown that the workload of VM versus the number of users served by VM mostly follows one of two profiles; constant or linear profiles as seen in Fig. 3. For example, in [10], the authors presented a CPU performance benchmark study for web application VMs serving a varying number of users with constant CPU workload as illustrated in Fig. 3 (a). Various benchmarking studies in the literature have demonstrated linear workload profiles for different applications with different slope coefficients. For example, Bharambe et al. presented a CPU performance benchmark study for multiple multiplayer games [11]. To maintain the service level agreement (SLA), each VM needs a minimum workload to run an application regardless of the number of users served by the VM, resulting in the workload profile shown in Fig. 3 (b). The minimum workload required to serve a user in a VM varies from as low as 1% to 60%.

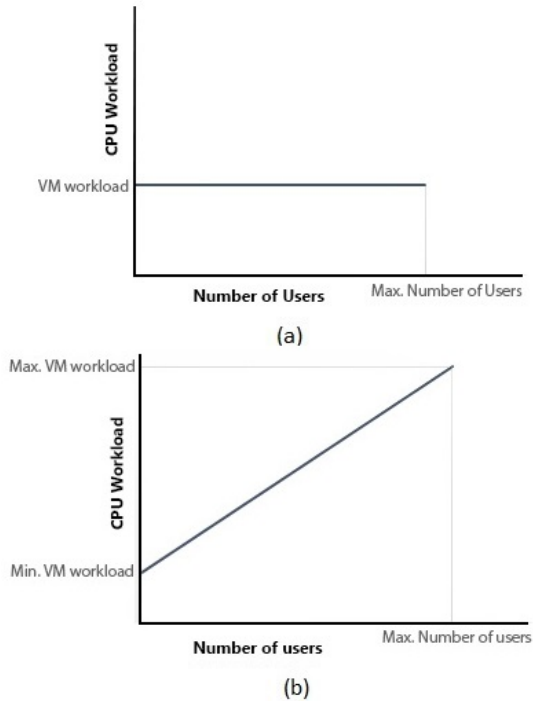

Figure 3: Relationship between VM workload and the number of users; (a) constant (b) linear relationship between VM workload and number of users.

From the network bandwidth perspective, according to Cisco Global Cloud Index: Forecast and Methodology 2016-2021 report [12], cloud and fog service applications based on user network bandwidth requirements are classified into three main categories: (i) basic applications (e.g. emails, web browsing, web-based learning system, Standard-Definition (SD) video applications etc), (ii) intermediate applications (web-based health records, augmented reality gaming applications, Voice over LTE, High-Definition (HD) video applications and related); and (iii) advanced applications (e.g. Ultra-HD (UHD) video applications). The data rate requirements for these three categories are typically: (i) up to 0.75 Mbps; (ii) between 0.75 Mbps and 2.5 Mbps and (iii) higher than 2.5 Mbps, respectively.

In a cloud environment, there are two types of inter-VM traffic; cooperation traffic and synchronization traffic. For cooperation traffic, in addition to sending traffic to users, each VM has another VM to cooperate with (e.g. traffic between an application VM and a database VM) as seen in Fig. 4(a). Synchronization traffic arises when, for example, VM replicas are created to serve distributed users, where these replicas need to be synchronized to each other (see Fig. 4(b)) to keep the content at each location up to date (e.g. social media where a user creates a post/webpage and posts it in the nearest VM replica, then, all other replicas need to be synchronized). VMs co-located in the same datacentre can communicate with each other through LAN, whereas, if VMs are located in geo-distributed datacenters, the communication traffic will pass through the core network backbone. The inter-VM traffic is a major contributor to the east-west traffic (server to server traffic) which is expected to be responsible for 85% of the global cloud traffic by 2021 as opposed to north-south (traffic between servers and users), which accounts for the remaining traffic [12].

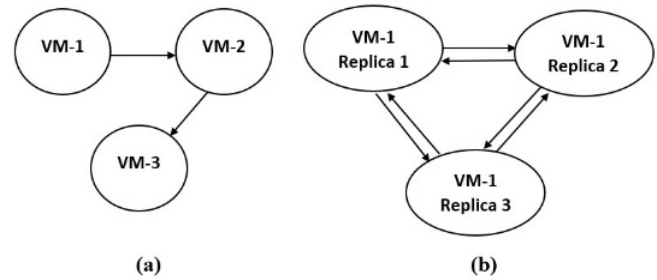

Figure 4: Illustrative example of inter-VM traffic, (a) VM-VM cooperation traffic (b) VM replicas synchronization.

4) *VMs Placement*

VMs placement is a vital process where the most suitable server is selected to host a VM. Selecting an appropriate VM container is critical to improving the energy efficiency of physical resources. However, VM placement in cloud computing was demonstrated to be a very complex task. The arrival patterns of simultaneous VM requests are usually unpredictable. Also, the datacenter size is typically enormous and for a certain load, optimizing the VMs placement is a nondeterministic polynomial (NP)-hard problem [5], [8].

VM placement schemes are classified as static and dynamic schemes. In the static schemes, the placement of the VM is fixed and is not adjusted for a long period of time whereas in the dynamic schemes, the VM placement can change periodically based on changes in the status i.e. server load, changes in the network or cloud components. Dynamic schemes can be categorized into reactive and proactive VM placements. In the reactive schemes, the placement of the VM is only changed if the system reaches undesired conditions. As such, the placement of the VM will only be changed for example in case of an emergency or when applying maintenance routines, restoring QoS or to reduce power consumption. On the other hand, in the proactive schemes, VM placement is changed before the system reaches undesired conditions.

**Energy Efficient VM Placement:** Under-utilized servers or poorly optimized VM placement can significantly increase the energy consumption, and consequently increase the carbon emissions and operating costs of cloud datacenters. In general, energy efficient VM placement algorithms should consider the following factors:

- Cloud and fog computing resources (servers, datacentre networks).
- Communication network layers (core, metro and access networks).
- VM-users and VM-VM traffic.
- QoS and SLAs.

The objective is to optimize the placement of VMs on a distributed cloud-fog architecture and to create an optimum number of replicas that results in minimum power consumption. Thus, designing energy efficient cloud-fog architectures requires the co-optimization of above-mentioned factors. The problem of providing energy-efficient VMs placement over end-to-end cloud-fog architecture considering the above-mentioned architecture has received little attention. For example, Misra et al. [13] built a theoretical model of fog computing and compared it with the conventional cloud computing model. In addition to the low latency, they found that offloading applications to fog nodes can significantly reduce power consumption by 41%. However, their investigation did not consider a detailed model of the telecom network architecture. The work in [14] considered energy efficiency in big data distributed cloud data centres, but these were located in the core nodes of the network only. Our work in [5] considered cloud fog architectures and their energy efficiency and as such this is an exception.

### III. NUMERICAL ANALYSIS

In this section, a comprehensive novel framework is developed based on mathematical mixed integer linear programming (MILP) modelling to study the offloading of VMs from the cloud to the fog layers with the objective of minimizing the total power consumption of providing the VMs. The placement of VMs in the cloud at the core network allows VMs to serve users distributed across the core nodes whereas placing the VM replicas closer to the users in the fog nodes in the metro or access network will ensure that the associated traffic between users and VMs does not traverse the core network and this therefore reduces the network power consumption. It will however increase the processing power consumption due to the creation of multiple replicas of the VMs. The creation of VM replicas therefore results in power savings if the former power consumption exceeds the latter power consumption. Overall, the power consumption can be reduced if the VM users traffic is high, inter-VM traffic is low, and/or the VMs have a linear power profile. In situations where such a linear profile exists, the creation of multiple VM replicas does not increase the power consumption significantly (there may be a slight increase due to idle / baseline power consumption) if the number of users remains constant.

The optimal VMs placement over AT&T cloud-fog architecture is investigated (AT&T core networks topology is illustrated in Fig. 5 [5]). We optimized a single VM as the simplest representative problem. This section investigates how the energy efficient placement of a single VM over a cloud-fog architecture varies based on three factors; the CPU requirements, download traffic and PUE values. The impact of the VM workload on the VM placement is examined by considering constant and linear workload profiles. The VM is considered to have 800 users. The workload of the constant workload profile VM; and the workload of the linear workload VM assume: 10%, 50% or 100% of the server CPU capacity in the scenarios studied. The VM with linear profile is considered to have no baseline (ie no idle power). The users are considered to access the VM with one of following download rates; 0.1 Mbps, 1 Mbps, 10 Mbps, 20 Mbps, 50 Mbps, 100 Mbps or 200 Mbps. Based on typical US datacentre energy usage [15], the power usage effectiveness (PUE) of the data centre varies based on the datacentre size. PUE is the ratio of the total data centre power consumption (including cooling and lighting) to the power consumption of IT and networking equipment. As more efficient cooling technologies are used in larger datacenters the PUE value approaches unity. For best practice datacenters, PUE of clouds, metro fog nodes and access fog nodes take the values 1.3, 1.4 and 1.5, receptively.

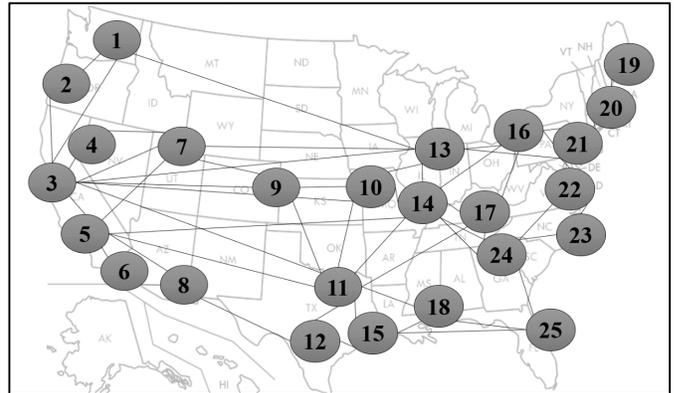

Figure 5: AT&T core network topology.

Figs. 6 (a), (b) and (c) show the optimal placement of VMs of 10%, 50% and 100% CPU requirements, respectively, considering best practice PUE values. In each figure, the x-axis is the VM workload profile, the y-axis is the data rates which range from 0.1 Mbps to 200 Mbps and the z-axis is the percentage of VM replicas in each location over the cloud-fog architecture. The placement of VMs with linear workload profile is not affected by the VM workload as, in this case, serving users will consume the same power whether centralized in a single VM or distributed among multiple replicas with smaller workloads. However, the higher PUE of fog nodes compared to the cloud, results in a situation where distributing replicas into fog processing nodes incurs additional power consumption as the PUE value of fog nodes is higher than clouds. Hence, there is a trade-off between network power saved by replicating VMs into fog nodes and the additional power consumed by these replicas. The creation of a VM replica results in power savings if the former power consumption exceeds the latter power consumption. At data rates of 1 Mbps and higher, VMs of 10%, 50% and 100% workloads are offloaded to access fog processing nodes considering a linear workload profile.

For constant workload profile, replicas are less energy efficient, therefore, offloading VMs to fog nodes decreases as the VM workload increases. While VMs of 10% workload and 20 Mbps are fully offloaded to metro fogs, 50% and 100% workload VMs are replicated only to clouds. Also, users of VM of 50% workload at 100 Mbps data rate as well as VMs of 100% workload at 200 Mbps data rate are served by clouds and metro fog nodes. A VM replica is offloaded to 14 metro fog nodes (in core nodes 1, 2, 4, 6, 7, 8, 13, 16, 19, 20, 21, 22, 23, 25) while users from other nodes are served by the replica placed in the cloud in core node 11 which they can access by traversing a single hop in the core network. These 14 metro fog nodes are selected to host replicas of the VM as the traffic flows will traverse more than a single hop in the IP over WDM network to access the VM placed in the cloud hosted in node 11 and therefore increase the needs for IP router ports (the most power consuming device in the IP over WDM network).

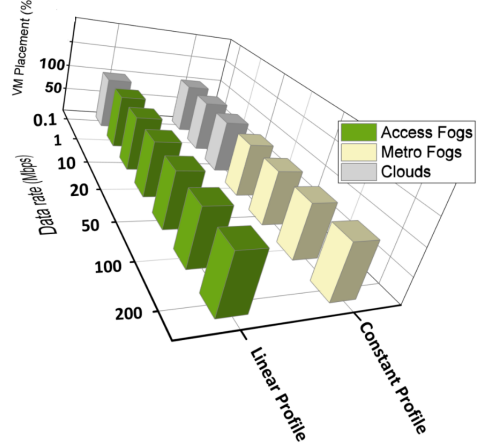

(a)

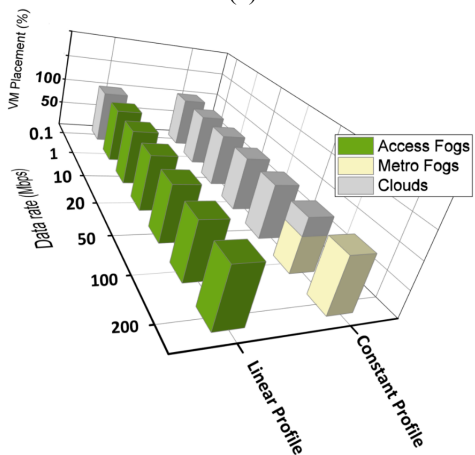

(b)

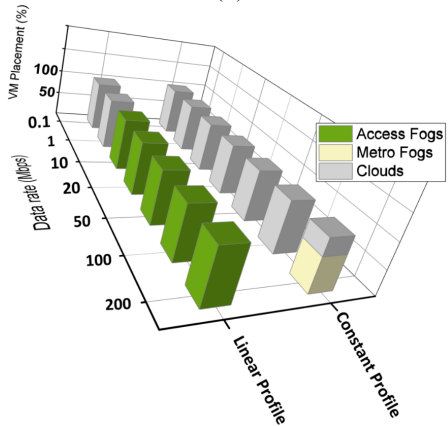

(c)

Figure 6: optimal VM placement of (a) constant profile at 10% of CPU and linear profile with peak utilization at (a) 10% case, (b) 50% case, (c) 100% case at different data rates considering best practice PUE value ($cloud\ PUE$=1.3, $Metro\ Fog\ PUE = 1.4$, $Access\ Fog\ PUE = 1.5$).

## IV. CONCLUSIONS

This article has tackled the problem of providing on demand services in an energy efficient manner over cloud-fog architectures considering the different communication and computation layers involved. It provided an overview of the communication networks used in cloud-fog service delivery focusing on IP over WDM networks, metro Ethernet networks and PON networks. These are the essential elements that represent the networks of interest in this article. Attention was then given to cloud computing and its recent introduction as an integral element of communication networks. The recent migration of processing and storage services from the cloud to the fog was outlined. A key enabling technology for the efficient utilization of cloud and fog resources is virtualization and the creation of virtual machines. These concepts were reviewed including VM categories, VM placement and energy efficient VM placement. Finally, the solution of the problem of energy efficient VMs placement explained using a range of numerical results.

**Acknowledgements**

The authors would like to acknowledge funding from the Engineering and Physical Sciences Research Council (EPSRC), INTERNET (EP/H040536/1), STAR (EP/K016873/1) and TOWS (EP/S016570/1) projects. The first author would like to acknowledge the Government of Saudi Arabia and Taibah University for funding his PhD scholarship. All data are provided in full in the results section of this paper.

**Biographies**

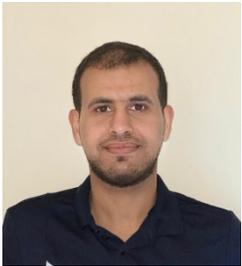

**Hatem A. Alharbi** received the B.Sc. degree in Computer Engineering (Hons.) from Umm Alqura University, Makkah, Saudi Arabia in 2012, the M.Sc. degree in Digital communication networks (with distinction) from University of Leeds, United Kingdom, in 2015. He is currently pursuing the Ph.D. degree with the School of Electronic and Electrical Engineering, University of Leeds, UK. He is currently a Lecturer in Computer Engineering department in the School of Computer Science and Engineering, University of Taibah, Saudi Arabia.

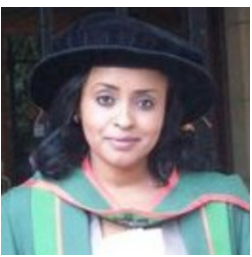

**Taisir EL-Gorashi** received the B.S. degree (first-class Hons.) in electrical and electronic engineering from the University of Khartoum, Khartoum, Sudan, in 2004, the M.Sc. degree (with distinction) in photonic and communication systems from the University of Wales, Swansea, UK, in 2005, and the PhD degree in optical networking from the University of Leeds, Leeds, UK, in 2010. She is currently a Lecturer in optical networks in the School of Electrical and Electronic Engineering, University of Leeds. Previously, she held a Postdoctoral Research post at the University of Leeds (2010– 2014), where she focused on the energy efficiency of optical networks investigating the use of renewable energy in core networks, green IP over WDM networks with datacenters, energy efficient physical topology design, energy efficiency of content distribution networks, distributed cloud computing, network virtualization and Big Data. In 2012, she was a BT Research Fellow, where she developed energy efficient hybrid wireless-optical broadband access networks and explored the dynamics of TV viewing behavior and program popularity. The energy efficiency techniques developed during her postdoctoral research contributed 3 out of the 8 carefully chosen core network energy efficiency improvement measures recommended by the GreenTouch consortium for every operator network worldwide. Her work led to several invited talks at GreenTouch, Bell Labs, Optical Network Design and Modelling conference, Optical Fiber Communications conference, International Conference on Computer Communications, EU Future Internet Assembly, IEEE Sustainable ICT Summit and IEEE 5G World Forum and collaboration with Nokia and Huawei.

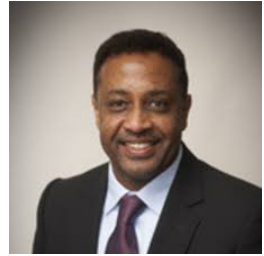

**Jaafar M. H. Elmirghani** is the Director of the Institute of Communication and Power Networks within the School of Electronic and Electrical Engineering, University of Leeds, UK. He joined Leeds in 2007 and prior to that (2000–2007) as chair in optical communications at the University of Wales Swansea he founded, developed and directed the Institute of Advanced Telecommunications and the Technium Digital (TD), a technology incubator/spin-off hub. He has provided outstanding leadership in a number of large research projects at the IAT and TD. He received the Ph.D. in the synchronization of optical systems and optical receiver design from the University of Huddersfield UK in 1994 and the DSc in Communication Systems and Networks from University of Leeds, UK, in 2014. He has co-authored Photonic switching Technology: Systems and Networks, (Wiley) and has published over 500 papers. He has research interests in optical systems and networks. Prof. Elmirghani is Fellow of the IET, Fellow of the Institute of Physics and Senior Member of IEEE. He was Chairman of IEEE Comsoc Transmission Access and Optical Systems technical committee and was Chairman of IEEE Comsoc Signal Processing and Communications Electronics technical committee, and an editor of IEEE Communications Magazine. He was founding Chair of the Advanced Signal Processing for Communication Symposium which started at IEEE GLOBECOM'99 and has continued since at every ICC and GLOBECOM. Prof. Elmirghani was also founding Chair of the first IEEE ICC/GLOBECOM optical symposium at GLOBECOM'00, the Future Photonic Network Technologies, Architectures and Protocols Symposium. He chaired this Symposium, which continues to date under different names. He was the founding chair of the first Green Track at ICC/GLOBECOM at GLOBECOM 2011, and is Chair of the IEEE Sustainable ICT Initiative within the IEEE Technical Activities Board (TAB) Future Directions Committee (FDC) and within the IEEE Communications Society, a pan IEEE Societies Initiative responsible for Green and Sustainable ICT activities across IEEE, 2012-present. He is and has been on the technical program committee of 38 IEEE ICC/GLOBECOM conferences between 1995 and 2019 including 18 times as Symposium Chair. He received the IEEE Communications Society Hal Sobol award, the IEEE Comsoc Chapter Achievement award for excellence in chapter activities (both in 2005), the University of Wales Swansea Outstanding Research Achievement Award, 2006, the IEEE Communications Society Signal Processing and Communication Electronics outstanding service award, 2009, a best paper award at IEEE ICC'2013, the IEEE Comsoc Transmission Access and Optical Systems outstanding Service award 2015 in recognition of "Leadership and Contributions to the Area of Green Communications", received the GreenTouch 1000x award in 2015 for "pioneering research contributions to the field of energy efficiency in telecommunications", the 2016 IET Optoelectronics Premium Award and shared with 6 GreenTouch innovators the 2016 Edison Award in the "Collective Disruption" Category for their work on the GreenMeter, an international competition, clear evidence of his seminal contributions to Green Communications which have a lasting impact on the environment (green) and society. He is currently an editor of: IET Optoelectronics, Journal of Optical Communications, IEEE Communications Surveys and Tutorials and IEEE Journal on Selected Areas in Communications series on Green Communications and Networking. He was Co-Chair of the GreenTouch Wired, Core and Access Networks Working Group, an adviser to the Commonwealth Scholarship Commission, member of the Royal Society International Joint Projects Panel and member of the Engineering and Physical Sciences Research Council (EPSRC) College. He was Principal Investigator (PI) of the £6m EPSRC INTelligent Energy awaRe NETworks (INTERNET) Programme Grant, 2010-2016 and is currently PI of the £6.6m EPSRC Terabit Bidirectional Multi-user Optical Wireless System (TOWS) for 6G LiFi Programme Grant, 2019-2024. He has been awarded in excess of £30 million in grants to date from EPSRC, the EU and industry and has held prestigious fellowships funded by the Royal Society and by BT. He was an IEEE Comsoc Distinguished Lecturer 2013-2016.